# The STONE curve: A ROC-derived model performance assessment tool

**Michael W. Liemohn,[1] Abigail R. Azari,[1] Natalia Yu. Ganushkina,[1,2] and Lutz Rastätter[3]**

[1]Department of Climate and Space Sciences and Engineering, University of Michigan, Ann Arbor, MI.

[2]Finnish Meteorological Institute, Helsinki, Finland

[3]Community Coordinated Modeling Center, NASA Goddard Space Flight Center, Greenbelt, MD

Corresponding author: Michael Liemohn (liemohn@umich.edu)

**Key Points:**

- A new event-detection-based metric for model performance appraisal is given with sliding thresholds in both observational and model values

- The new metric is like the relative operating characteristic curve but uses continuous observational values, not just categorical status

- The new metric is used on real-time model predictions of common geomagnetic activity parameters, demonstrating its features and strengths

**AGU Index Terms:**

- 1984         Statistical methods: Descriptive (4318)
- 4318         Statistical analysis (1984, 1986)
- 7924         Forecasting (1922, 2722, 4315)
- 0550         Model verification and validation
- 9820         Techniques applicable in three or more fields

## Abstract

A new model validation and performance assessment tool is introduced, the sliding threshold of observation for numeric evaluation (STONE) curve. It is based on the relative operating characteristic (ROC) curve technique, but instead of sorting all observations in a categorical classification, the STONE tool uses the continuous nature of the observations. Rather than defining events in the observations and then sliding the threshold only in the classifier/model data set, the threshold is changed simultaneously for both the observational and model values, with the same threshold value for both data and model. This is only possible if the observations are continuous and the model output is in the same units and scale as the observations, i.e., the model is trying to exactly reproduce the data. The STONE curve has several similarities with the ROC curve – plotting probability of detection against probability of false detection, ranging from the (1,1) corner for low thresholds to the (0,0) corner for high thresholds, and values above the zero-intercept unity-slope line indicating better than random predictive ability. The main difference is that the STONE curve can be nonmonotonic, doubling back in both the x and y directions. These ripples reveal asymmetries in the data-model value pairs. This new technique is applied to modeling output of a common geomagnetic activity index as well as energetic electron fluxes in the Earth's inner magnetosphere. It is not limited to space physics applications but can be used for any scientific or engineering field where numerical models are used to reproduce observations.

## Plain Language Summary

Scientists often try to reproduce observations with a model, helping them explain the observations by adjusting known and controllable features within the model. They then use a large variety of metrics for assessing the ability of a model to reproduce the observations. One such metric is called the relative operating characteristic (ROC) curve, a tool that assesses a model's ability to predict events within the data. The ROC curve is made by sliding the event-definition threshold in the model output, calculating certain metrics and making a graph of the results. Here, a new model assessment tool is introduced, called the sliding threshold of observation for numeric evaluation (STONE) curve. The STONE curve is created by sliding the event definition threshold not only for the model output but also simultaneously for the data values. This is applicable when the model output is trying to reproduce the exact values of a particular data set. While the ROC curve is still a highly valuable tool for optimizing the prediction of known and pre-classified events, it is argued here that the STONE curve is better for assessing model prediction of a continuous-valued data set.

## 1. Introduction

Numerical models are a fundamental feature of research in the natural sciences. Models are often used to explain strange and interesting features in an archival data set in order to assess the physical processes responsible for that observational signature. They are also used for prediction, using some estimate of future initial and boundary conditions to determine the state of the system, or even a particular observational quantity, ahead of time. These are typical uses of models in every discipline of Earth and space sciences.

There exists a large collection of metrics to assess the goodness of fit for these models to a particular data set. These metrics, for the most part, can be sorted into several major groupings, two of which are fit performance metrics and event detection metrics (e.g., Wilks, 2011; Joliffe and Stephenson et al., 2012; Liemohn, McCollough, et al., 2018). The former group, also called continuous metrics, is based on differencing each data-model value pair and includes many well-known assessment equations such as root mean square error, correlation coefficient, mean error, and prediction efficiency (e.g., Hogan and Mason, 2012; Morley et al., 2018). The second group, also called categorical metrics, is based on categorizing the observations into events and non-events and then assessing a model's ability to reproduce this classification. This is done through a contingency table (also commonly called a confusion matrix) in which each data-model pair gets two designations: determining if the observation is in the event state or not and similarly if the model value is in the event state or not. The similarity or difference of the data and model values is irrelevant, only the event/non-event designation matters. This second group includes other well-known assessment equations such as the probability of detection, false alarm rate, frequency bias, and Heidke skill score (see, e.g., Muller et al., 1944; Wilks, 2011).

A feature of the event detection metrics is that the model does not have to cover the same range or even have the same units as the observations. The model could be anything that might predict the event state of the observations. Furthermore, the observations do not have to be a continuous-valued real number set, but could be pre-categorized into events and non-events (or a multi-level classification). The model could be a continuous-valued real number set or a discrete-valued categorized set. When the data or model happens to be a continuous-valued real number set, then a threshold value for event identification is chosen, a threshold value that could be different between the observational events and the modeled events.

An event detection metric that is often used for weather prediction (e.g., Mason, 1982), psychology (e.g., Swets, 1972), medical clinical trials (e.g., Ekelund, 2011), and machine learning (e.g., Fawcett et al., 2006) is the relative (or, originally, receiver) operating characteristic (ROC) curve (see review by Carter et al., 2016). This is an assessment tool that can be applied when the model values are continuous-valued real numbers, using not just one event identification threshold but many. The method is to sweep the event definition threshold for the model values from low to high, calculating two specific metrics, the probability of detection (POD) and the probability of false detection (POFD), and plotting these two arrays against each other. The threshold that yields the location on the ROC curve closest to the upper left corner (high POD and low POFD) can be considered a possible "best setting" for event prediction by this model. This is not the only location for an optimum pick of a final threshold along an ROC curve. Often the final choice will depend on the application and problem specific details. For example, recent developments have discussed the use of skill scores for different solar and space physics applications (e.g., Bobra & Couvidat 2015) and their location on ROC diagrams (e.g., Manzato, 2007; Azari et al., 2018). A further detailed discussion on skill scores and their relation to ROC diagrams can be found within Manzato (2005). An integral quantity sometimes used from the ROC curve is the area under the curve (AUC), which is an overall measure of goodness of fit for the model to the observational events across all of the possible model value event identification thresholds.

The ROC curve has recently been used quite often in the Earth and space sciences to assess model performance at detecting events in an observational data set. It is used regularly in the atmospheric sciences, such as for regional ozone ensemble forecasting (e.g., Delle Monache et al., 2006), deciphering the microphysical properties of clouds (e.g., Gabriel et al., 2009), and forecasting summer monsoons over India (e.g., Borah et al., 2013). Earth scientists also employ the ROC curve for a diverse set of modeling activities, including the distribution of rock glaciers (e.g., Brenning et al., 2007), assessing triggering mechanisms of earthquake aftershocks (e.g., Meade et al., 2017), and snow slab instability physics (e.g., Reuter & Schweizer, 2018). This also includes land-air interactions, such as mapping of expected ash cloud locations after eruptions (e.g., Stefanescu et al., 2014), modeling rainfall-induced landslides (e.g., Anagnostopoulos et al., 2015), and statistically forecasting extreme corn losses in the eastern United States (Mathieu & Aires, 2018). The fields of space and planetary science have also started to employ this technique, such as for oblique ionogram retieval algorithm assessment (Ippolito et al., 2016), identifying energetic particle flux injections at Saturn (e.g., Azari et al., 2018), magnetic activity prediction (e.g., Liemohn, McCollough, et al., 2018), and identifying solar flare precursors (e.g., Chen et al., 2019). In short, the ROC curve has become an essential tool, among many that can and should be applied, for model assessment across many natural science disciplines.

The ROC curve, however, only assesses the model's ability to predict a single observational event identification threshold. While this is desirable if the data were pre-classified as events or non-events, this imposes a simplification of the data set when the observations are also continuous-valued real numbers. That is, the ROC curve does not test the model's ability to predict events across the full range of the data. A family of ROC curves can be produced using different data-value event identification thresholds (and sweeping the model-value event identification threshold to produce each ROC curve), which is acceptable if the model is only being used to maximize the prediction of events. If the model, however, is trying to reproduce the exact values of the observations, then it is useful to conduct an assessment for which the data and model have the same threshold setting. The ROC curve, unfortunately, cannot easily test the model's ability to reproduce the observed events at the same threshold setting, sweeping through all possible event identification thresholds.

There exists a need for a new metric. Like the ROC curve, this new metric should test a model's ability to predict observed events across the full range of possible model-value event identification settings, but rather than using a single observational event categorization, it should sweep through the same range of event identification thresholds as used for the model. Such a metric is proposed below, called the sliding threshold of observation numeric evaluation, or STONE, curve. This is based on the ROC curve but includes the desirable features described above. The work then presents an application of the STONE curve to two space physics data sets, the prediction of a geomagnetic activity index and energetic electron fluxes in near-Earth space. Similarities and differences between the ROC and STONE curves are discussed, as well as the interpretive meaning of features in the STONE curve.

## 2. Method of Calculation

The calculation of a STONE curve is rather similar to that of a ROC curve, with one major exception – both thresholds slide together, incrementing the two event identification thresholds simultaneously so that the same threshold value is used for both the data and the model at each setting from low to high across the range. Because this tool is for continuous-valued observations and model results, for which an "event" is an arbitrary designation, there does not have to be a pre-defined event threshold in the observations. In fact, it is desired that the model match the observations for all levels of "event" definition. Therefore, in the STONE tool, the two thresholds move together. This is illustrated in Figure 1, showing an arbitrary data set plotted against a model output that is trying to reproduce these values.

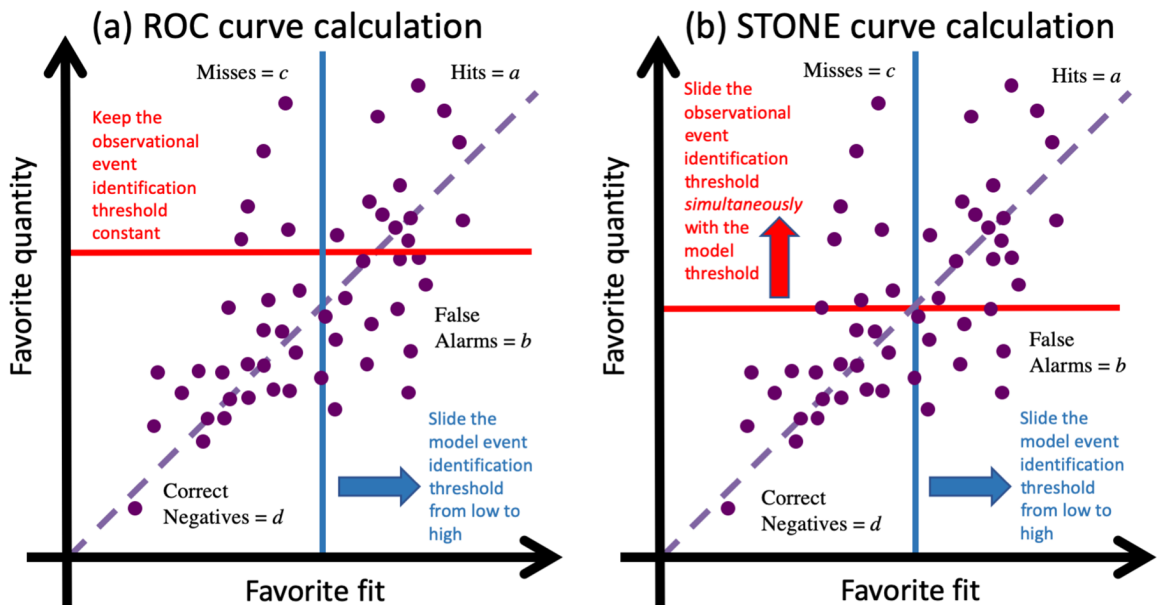

**Figure 1.** Idealized examples of how to calculate (a) the ROC curve and (b) the STONE curve. In (a), only the blue curve shifts while the red curve is fixed at some level. In (b), both the red and blue thresholds shift together. As these lines shift, data points are converted from one quadrant to another. The purple dashed curve is the zero-intercept unity-slope line, for reference.

Figure 1a shows the calculation scenario for the ROC curve, with the event identification threshold for the observations set to a fixed value and the threshold for the model results sweeping from low to high values. Annotations label the four quadrants of the chart, as defined by these two thresholds. As the model threshold changes, the points in the chart change quadrant. Specifically, two shifts occur: points in the "hits" quadrant (variable *a*) move to the "misses" quadrant (*c*) and points in the "false alarms" quadrant (*b*) move to the "correct negatives" quadrant (*d*).

The ROC curve is defined from two metrics in the "discrimination" category (Murphy & Winkler, 1987) of data-model comparison techniques. Discrimination metrics are assessments that only use a portion of the data values within a specified range (and the corresponding model values). For event detection metrics, the usual practice is to use the event state of the observations to define the subsets of the data. In particular, the ROC curve uses POD and POFD, which have the following formulas:

$$POD = \frac{a}{a+c} \tag{1}$$

$$POFD = \frac{b}{b+d} \tag{2}$$

Where *a*, *b*, *c*, and *d* are point counts from the quadrants in the scatter plot. It is seen that these two formulas are mutually exclusive, POD only uses the hits and misses quadrants while POFD only uses the false alarms and correct negatives quadrants. Because the data threshold remains fixed for the ROC curve, the points either contribute to POD or POFD, regardless of the model threshold designation. For a very low model threshold setting, all of the points are in either the hits or false alarms quadrants, which sets both POD and POFD to one. As the model threshold is increased, points are converted from hits to misses and from false alarms to correct negatives, which monotonically decreases POD and POFD. For a very high model threshold, all of the points will then be misses or correct negatives, and both POD and POFD will be zero.

Figure 1b shows the calculation scenario for the STONE curve. In this situation, both event identification thresholds move simultaneously. The four quadrants are still defined as with the ROC curve, but with both thresholds changing, the shift of points from one quadrant to another is not so simple. For a very low threshold setting, nearly all points will be hits and perhaps a few will be false alarms. Thus, like the ROC curve, the STONE curve also begins in the (1,1) corner of POFD-POD space (assuming a "low" starting threshold value). Also similarly, for a very large threshold setting, nearly all points will be correct negatives and perhaps a few will be misses, with the STONE curve ending in the (0,0) corner of POFD-POD space. Another similarity is that false alarms are converted into correct negatives as the threshold setting increases.

The big difference between the ROC and STONE curve calculations, however, is that as the event identification threshold increases, a hit event can shift to any of the other three quadrants. If it is far above the data threshold but close to the model threshold, then the threshold increase will cause the point to shift from being a hit to a miss. If it is close to the data threshold but far away from the model threshold, then it will shift from being a hit to being a false alarm. If it is close to both thresholds, then there is a chance it will cross both lines during the incremental shift and jump from the hits regions to the correct negatives zone. Only the first of these three moves (hits to misses) occurs with the ROC curve calculation. In addition, misses are shifting to become correct negatives as the observational threshold is incremented to higher values, another move that is not part of the ROC curve calculation. The behavior of the POD and POFD values as a function of threshold, therefore, are not intuitively known and the STONE curve does not have to be monotonic between its (1,1) and (0,0) endpoints.

## 3. Application of the STONE tool

With this definition for the STONE curve, it can be used on a few example data-model comparisons to illustrate the similarities and differences with the ROC curve. Here, two comparisons will be shown. The first is for a model prediction of a geomagnetic activity index, originally presented by Liemohn, Ganushkina, et al. (2018), and the second is for energetic electrons in near-Earth space, originally presented by Ganushkina et al. (2019).

### 3.1. Predicting a geomagnetic activity index

Liemohn, Ganushkina, et al. (2018) compared the output from experimental real-time simulations of the Space Weather Modeling Framework (SWMF) against the disturbance storm-time index, Dst (Rostoker et al., 1972). The SWMF is a collection of space physics numerical models simulating the Sun-Earth space environment (Toth et al., 2012), and in many other planetary environments (e.g., Jia et al., 2012; Ma et al., 2013; Dong et al., 2014; Liemohn et al., 2017). This geospace environment simulation has a very similar setup to that of Pulkkinen et al. (2013), using the Block Adaptive Tree Roe-type Upwind Scheme (BATS-R-US) magnetohydrodynamic model coupled to the Rice Convection Model (RCM) and the Ridley Ionosphere Model (RIM). Real-time solar wind and interplanetary magnetic field input was taken from the Advanced Composition Explorer (ACE) satellite. The simulated Dst time series from the SWMF was calculated with the method from Yu et al. (2010) and compared against the real-time version of the Dst index as produced by the Kyoto World Data Center for Geomagnetism. The interval of comparison spans from 19 April 2015 until 17 July 2017, which is 27 months of 1-hour resolution measurements and corresponding model output values (just under 300,000 data-model pairs).

Figure 2a shows a scatter plot of the SWMF Dst values against the observed Dst values. While the individual points are analyzed as unique contributions, they are binned to produce the colored curves on the plot, demarking contours of 50 points within a 5-by-5 nT grid. Note that, because Dst is near zero for quiet times and shifts to negative values during storm times, events are defined as values below (i.e., more negative) a chosen threshold. As defined by Gonzalez et al. (1994), a typical designation for the Dst index measuring a storm situation is -30 nT or below for a weak storm and -50 nT or below for a moderate storm, so these two settings are used for the ROC curve observational threshold setting. These two thresholds are indicated in Figure 2a as horizontal dashed lines.

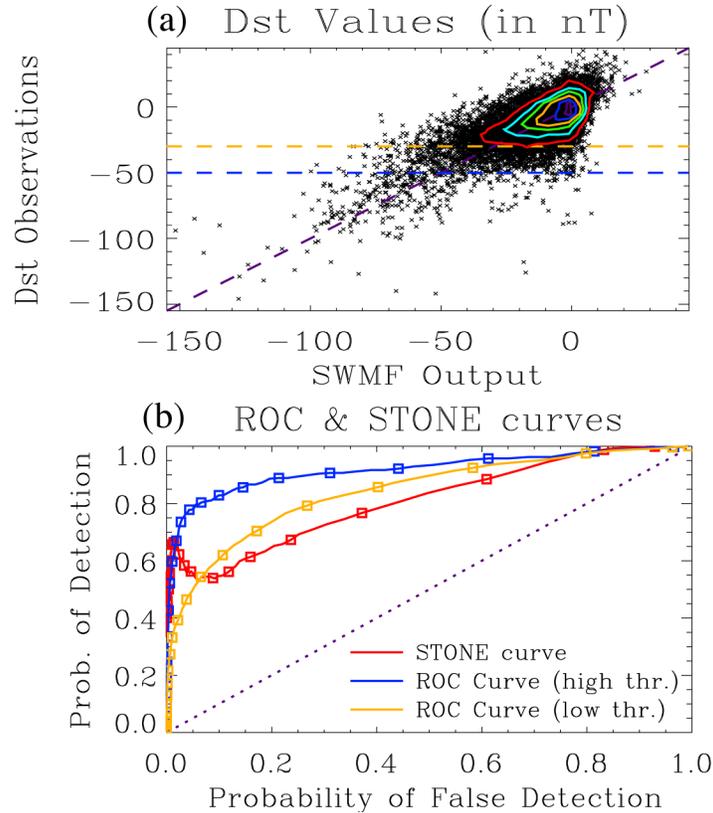

**Figure 2.** (a) Scatter plot of the observed real-time Dst time series (y-axis values) against a prediction Dst time series from the SWMF (x-axis values). The contours are drawn every 50 points per 5x5 nT bin. Also drawn are horizontal dashed lines at the ROC event thresholds of -30 and -50 nT, with events defined as the points below these lines. A purple dashed zero intercept unity slope line is also drawn, for reference. (b) STONE (red) and ROC curves (blue for -50 nT, orange for -30 nT observed event threshold) calculated from the scatter plot. Symbols are shown along all three curves at every 5 nT threshold increment. The diagonal dotted line with zero intercept and unity slope is shown for reference.

The ROC and STONE curves are calculated as follows and shown in Figure 2b. To create a ROC curve, the model threshold setting is initially set to +10 nT and then swept in 1 nT increments to -120 nT. The data threshold for events is held fixed, at -50 nT for the blue curve and -30 for the orange curve. To create STONE curve (red line), this same model threshold variation is followed, but the data threshold is also swept from +10 to -120 nT. Symbols are shown along each of the plots every 5 nT of threshold increment.

Some features of Figure 2b should be noted. It is seen that the ROC curves monotonically increase from (0,0) to (1,1). The ROC curve with a -50 nT event threshold is well above the zero-intercept, unity-slope line (the diagonal purple dotted line on Figure 2b), indicating that the model is reasonably good at reproducing moderate and stronger storm events recorded by the real-time Dst index. The closest approach to the upper left corner occurs at a threshold of -37 nT for the -50 nT threshold ROC curve and -17 nT for the -30 nT ROC curve, which indicates that the model somewhat underpredicts the strength of such storms.

The STONE curve lies both above and below these two ROC curves, depending on the threshold. The STONE curve is coincident with each ROC curve at the locations where the ROC curve model threshold setting is equal to the observational threshold setting (-30 nT for the orange curve, -50 nT for the blue curve). They cross elsewhere, too, such as in the low-threshold (i.e., a threshold of near and above zero) region in the upper right region of the plot. It is seen that the STONE curve is not monotonic but includes a local maximum and local minimum at the "high threshold" settings (minimum at -28 nT threshold and maximum at -52 nT threshold). The nonmonotonicity is because POD increases at these threshold values. An increase in POD is achieved by more points leaving the misses quadrant than leaving from the hits quadrant.

This is better understood by considering the distribution of points beyond a few threshold choices. Figure 3 shows histograms of the points above a particular data or model threshold setting. In particular, three threshold settings are displayed – -30 nT, -40 nT, and -50 nT – showing the points at "higher" (more negative) Dst values in both the data and model (left and right columns, respectively). For Figure 3a, the counts are for all points below some horizontal line of an event identification threshold setting of the observations. For Figure 3b, the counts are for all points to the left of some event identification threshold setting for the model values. The calculated skew for these distributions is listed in each panel.

In Figure 3a, it is evident, both qualitatively from the histograms and quantitatively from the skew values, that the distribution of model output values is significantly changing across these three observational threshold settings. For the more negative threshold, there are far fewer model values between zero and -50 nT. That is, across these threshold settings, many of the points in the misses quadrant were converted into correct negatives. In Figure 3b, the three distributions have essentially the same shape, with a large negative skew. These distributions do not undergo the same systematic alteration in their shape the way that the distributions in Figure 3a did. Putting these two features together, it means that more misses were removed than hits, and so POD increased as the STONE threshold was swept to more negative values between -30 nT and -50 nT. This resulted in a nonmonotonic wiggle in the STONE curve at these thresholds.

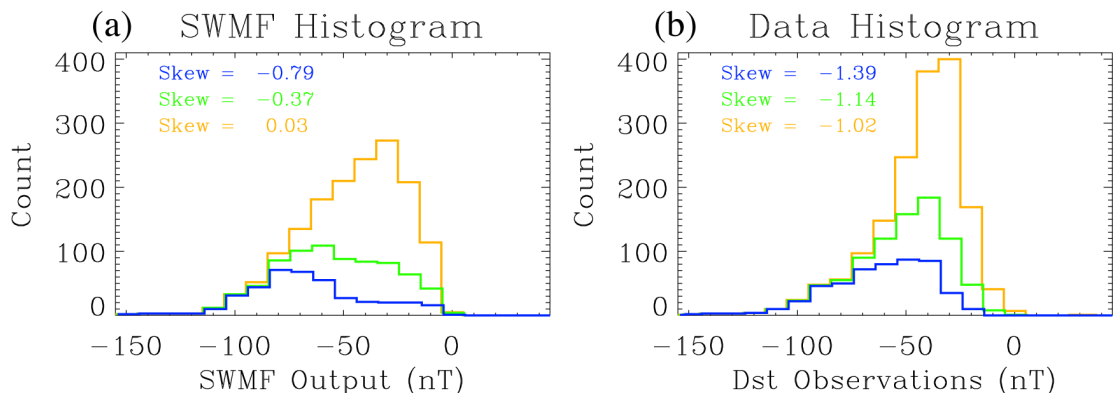

**Figure 3.** (a) Histogram of model values for all data values below three different thresholds: -30 nT (orange curve), -40 nT (green curve), and -50 nT (blue curve). (b) Histogram of data values for all model values below the same three thresholds. The bin sizes for each histogram is 10 nT. The calculated skew for each distribution is listed in each plot.

### 3.2. Predicting energetic electrons in near-Earth space

Ganushkina et al. (2019) compared real-time output from the inner magnetosphere particle transport and acceleration model (IMPTAM) with measurements from the magnetosphere electron detector (MAGED) on the geosynchronus orbiting environmental satellites (GOES) in geostationary orbit at 6.62 Earth radii geocentric distance over the American sector (Rowland & Weigel, 2012; Sillanpaa et al., 2017), specifically, with data from GOES-13, -14, and -15. IMPTAM, initially developed by Ganushkina et al. (2001) and used regularly for investigating the physics of plasma sheet electron transport (e.g., Ganushkina et al., 2013, 2014), has been running in a real-time operational mode since February 2013, first in Europe and then a mirror site at the University of Michigan. Ganushkina et al. (2015) made an initial comparison of these model output values against a few months of GOES data, while Ganushkina et al. (2019) provided a far more robust validation analysis of the model, covering over 18 months (September 20, 2013 through March 31, 2015). It is this second interval that will be used again for this study.

Figures 4a and 4b show two scatter plots comparing the IMPTAM and GOES electron differential number fluxes at 40 keV. The colored contours show the point density, with a new curve every 50 points within a bin (defined, for these contours, with 10 bins per decade in both the data and model values). Figure 4a presents the full data set while Figure 4b only shows the comparison for those values in the 03 to 09 magnetic local time (MLT) range, the region found by Ganushkina et al. (2019) to have a "good comparison" between the data and model values. On each of these plots, two observational event thresholds are shown as the horizontal dashed lines, drawn at $5 \times 10^4$ and $2 \times 10^5$ electrons cm$^{-2}$ s$^{-1}$ sr$^{-1}$ keV$^{-1}$ in green and blue, respectively.

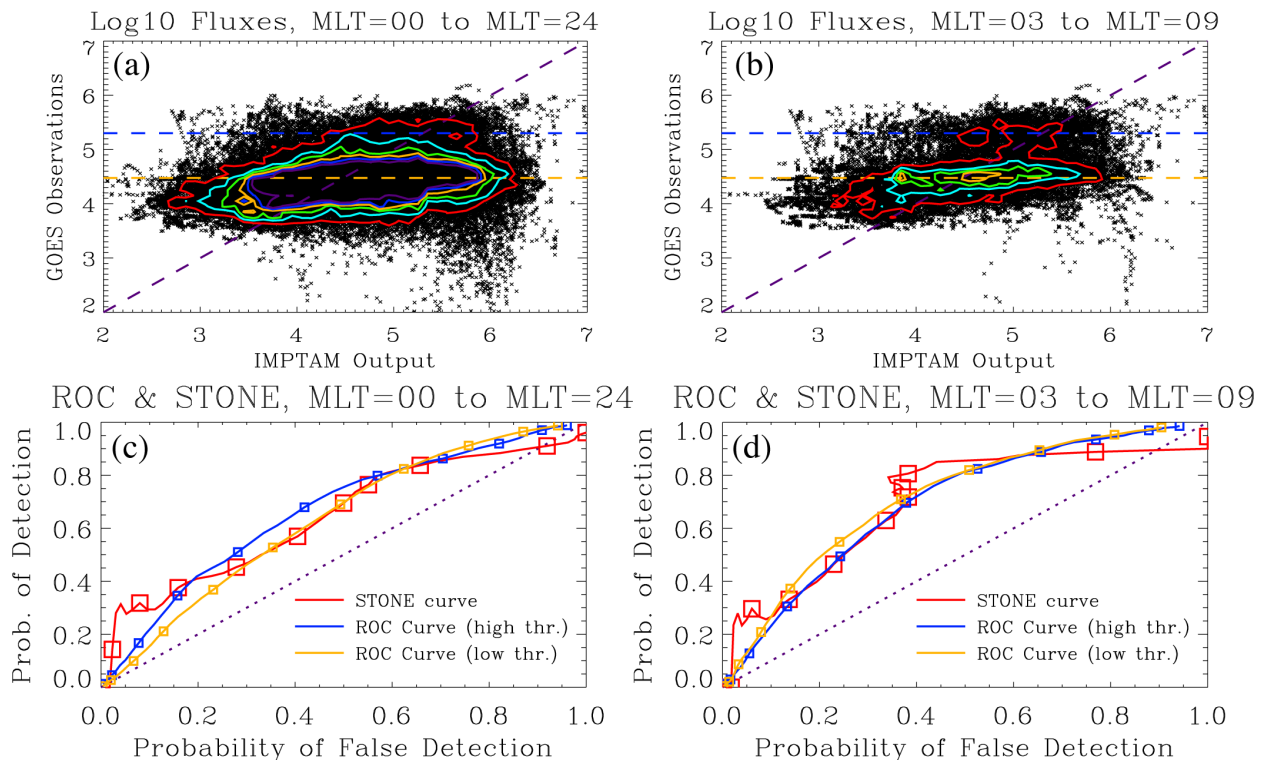

**Figure 4.** Scatter plot comparing GOES and IMPTAM 40 keV electron differential number fluxes (log base 10 of electrons $cm^{-2}$ $s^{-1}$ $sr^{-1}$ $keV^{-1}$) for (a) all MLTs and (b) the 03-09 MLT range. Color contours are shown every 50 points per bin (10 bins per decade in both data and model). The horizontal dashed lines show the ROC thresholds of $3x10^4$ and $2x10^5$. A purple dashed zero intercept and unity slope line is shown for reference. The lower panels show STONE curves (red) and ROC curves (blue for $2x10^5$ and orange for $3x10^4$) for (c) the full MLT comparison and (d) the 03-09 MLT range. Symbols are shown every factor of 2 increase in threshold value. The diagonal dotted line with zero intercept and unity slope is shown for reference.

Figures 4c and 4d show the ROC and STONE curves for these two data-model comparisons, the full set with values at all MLTs and the subset from 03 to 09 MLT, respectively. In both Figures 4c and 4d, the STONE curve again has a nonmonotonic shape at high threshold settings (above $4x10^5$). Like the similar case for the Dst STONE curve in Figure 2b, this shows that, for these thresholds, more points are being removed from the misses quadrant than being removed from the hits quadrant.

Figure 4d has another unusual feature in the STONE curve, seen as a nonmonotonicity in the x-axis values. This is from the POFD values increasing with increasing threshold (rather than decreasing, as they always do with a ROC curve). This is occurring for thresholds between $1x10^4$ and $4x10^4$, just as the STONE curve crosses the orange ROC curve. Considering equation (2) above, the correct negatives in the denominator are always increasing with increasing threshold, as points convert to this quadrant from any of the other three quadrants. For POFD to increase, the false alarms had to increase faster than the correct negative point count. This is seen in Figure 4b as the points have a horizontal peak (highlighted by the flat, elongated color contours). Many points are being converted from the hits quadrant into the false alarms quadrant and, for these threshold settings, this conversion to false alarms outpaces the conversion of points into the correct negatives quadrant. This results in a ripple in the STONE curve at these thresholds.

Figures 4c and 4d, the STONE curve is quite close to the two ROC curves, which are very similar to each other. This can be understood from the "flatness" of the cloud of points in the scatter plots in Figures 4a and 4b. The points are not well aligned with the zero intercept and unity slope line, revealing less than perfect agreement between the observations and model output. However, in terms of physics-based real-time modeling of near-Earth magnetospheric electron fluxes, this is actually quite good, arguably the best that is currently available. This means that all ROC curves will be close to each other, as any observational event identification threshold will have a relatively similar transfer of points between the quadrants. However, because the model is trying to exactly reproduce the observed flux values, the STONE curve can be calculated, and this new curve includes several nonmonotonicities. The wiggles and ripples in the STONE curve reveal thresholds where the distribution of points, in either the vertical or horizontal direction, are asymmetric, bi-modal, or otherwise non-Gaussian. The ROC curves cannot reveal this kind of information about the distribution of points in the scatter plot the way that the STONE curve can.

## 4. Discussion

The STONE curve introduced above is a new tool for assessing the ability of a model with a continuous-valued output to exactly match a continuous-valued data set. As illustrative example usages, it was applied to two recently-published data-model comparisons, a prediction of the disturbance storm-time index Dst and a prediction of energetic electron fluxes in near-Earth space.

The STONE curve is quite similar to the ROC curve. It is based on the same contingency table calculations of POD and POFD, plotting these two values against each other for a range of event threshold settings. Like the ROC curve, it starts at (1,1) for low threshold settings and moves to (0,0) for high threshold settings. Also like the ROC curve, being above the zero-intercept, unity-slope line indicates a prediction that is better than random chance. Curves are better when they are closer to the upper left corner in POFD-POD space, and a common choice for the best optimization point along along a ROC or STONE curve is that closest to this corner as this point reveals the best model threshold setting for optimizing discrimination performance. That is, both curves reveal a possible best model threshold setting for event prediction, the ROC curve revealing the best settings for a specified observational event identification threshold and the STONE curve revealing the best setting against the an identically defined observational event. Of course, this is "best" only if discrimination is what should be optimized for the particular application. A different threshold settings might be most favorable if other considerations outweigh discrimination, such as minimizing false alarms or maximizing a particular skill score.

Another similarity is that the integral area of the ROC curve, AUC, is equally applicable to the STONE curve. AUC, a synthesis of the entire threshold-setting range into a single number, indicates the quality of the chosen model to predict the events identified in the observational data (see the detailed explanation of AUC in Fawcett (2006) or Ekelund (2011)). Being an integrated quantity, AUC is a complementary metric to the "best model threshold setting for event prediction" mentioned in the preceding paragraph because AUC uses information from all model threshold settings, even those with POFD-POD coordinates far from the "best setting" upper-left corner of the graph. Comparing AUCs for several STONE curves (i.e., using different models against the same data set) will provide a quantitative assessment of which of the models has the best system-level predictive capability against that data set. It could be that the model with the highest AUC is not the model with a point along its STONE curve closest to (0,1) in POFD-POD space. Such a case reveals that the first model, with the higher AUC, has the best model physics for reproducing the data set as a whole, but that the second model is actually best at predicting events with a particular threshold setting. Because it is calculated the same way, AUC can be used to compare STONE curves just like it is for ROC curves.

A key difference between the STONE and ROC curves is that the STONE curve can have nonmonotonicities. These features, which can be wiggles with respect to either POD or POFD, reveal features of the model prediction of events that are not easily extracted from a ROC curve. This makes the STONE curve somewhat like a fit performance metric, even though it is an event-detection metric that disregards the difference between the data-model pairs.

The nonmonotonicities in the STONE curve reveal information about the distribution of points in the data-model comparison. Specifically, they show the existence of an asymmetry, perhaps a non-Gaussian point spread like a skewed or bimodal distribution, for the pairs above that threshold setting. Combined with a histogram or even fit-performance data-model comparison formulas for this subset of either the data or model values, the nature of this distribution can be explored.

Why not just start out by calculating fit performance metrics on these subsets? The answer is because the subset of interest would not have been known; the STONE curve revealed the thresholds where the distribution had a changing or non-Gaussian distribution. That is, it could be used to optimize the fit performance analysis by identifying the subset of the data or model that should be considered in more detail. Also, the STONE curve includes information not just within a subset of the data (discrimination) or a subset of the model (reliability), but includes information about the entire data-model comparison set, because POD and POFD use all data-model pairs in the point counting in the quadrants. For one of the specific examples in the manuscript: continuous metrics will tell the user very little about the SWMF's ability to predict magnetic storms of -50 nT or less. A ROC curve is far more suited to this, and a STONE curve one step farther, revealing the ability of the model to predict Dst levels below any threshold (which could be accomplished by a large family of ROC curves). No continuous metric that does this type of assessment. If the detection of events is desired, then the STONE curve is an advantageous assessment tool in addition to standard continuous fit performance metrics.

A useful follow-on study to this would be a detailed analysis of the features of the STONE curve to the underlying distribution of points in the data-model scatter plot. That is, by assuming known two-dimensional distributions of points of several different shapes and parameter settings, the connection between the distribution and the resulting features in the STONE curve can be isolated. Such an in-depth assessment of the STONE curve is beyond this initial description and illustrative usage of this metrics tool and is left as a future project.

A key feature of the STONE curve is that it reveals the threshold (or range of thresholds) for which the model does best at reproducing similarly-defined events in the data. A single ROC curve cannot do this because it uses a fixed threshold for identifying events in the observations. When the data are continuously-values and the model is seeking to reproduce these exact values, then it is useful to examine the event detection capability of the model at the same threshold settings between data and model. A single ROC curve doesn't do this, except at one threshold setting. The STONE curve, therefore, is a better assessment tool for models that are trying to predict the exact value of a data set.

The ROC is still a highly useful tool for event prediction and this study does not seek to replace it with the STONE curve. Indeed, the ROC curve is optimal for categorical data sets where the observations have been pre-classified as events and non-events. In this case, the STONE curve cannot be used because the data and model are on different scales, the former being a binary yes-no designation and the latter being either a real number range or its own categorical designation. The ROC curve can handle this difference in units while the STONE curve cannot.

The two example data-model comparisons to which the STONE curve was applied are both from space physics. The first was an evaluation between a physics-based model of geospace, running in real time, with the real-time version of the Dst index, a measure of geospace activity (see its comparison with other similar indices in Katus & Liemohn (2013)). Many models exist for the prediction of Dst (see the review by Liemohn, McCollough, et al., 2018), with some models doing exceptionally well at reproducing the observed time series.

While this chosen model for this comparison is arguably the best physics-based model for reproducing Dst (see, for comparison, the solar cycle storm-interval Dst comparison of Liemohn & Jakowski (2008)), it is not the best model available at predicting this index. In fact, many empirical models are substantially better at capturing the storm intervals of Dst. The second example was a comparison of a physics-based model of energetic electron fluxes in the near-Earth magnetosphere, running in real time, with real-time observations from a geosynchronous spacecraft. Magnetospheric charged particle fluxes are notoriously difficult to reproduce with physics-based modeling approaches (see, e.g., Morley et al., 2018), and even empirical models reduce the problem to remove the fast temporal dynamics, averaging over a day (e.g., Li, 2004) or an hour (e.g., Boynton et al., 2019). That is, these two examples represent state-of-the-art physics-based approaches to space weather nowcasting, but are not the best predictions of these two quantities across the field.

It is worth stating here that there are many other metrics in existence for evaluating a scatter plot of data-model values like that shown in Figure 1. No one metric equation or technique does everything; each was designed to assess only a specific aspect of the relationship. That is, neither the ROC curve nor the STONE curve should be used as the sole assessment tool for a model against a particular data set. In practice, many metrics, from both the continuous fit-performance grouping and from the categorical event-detection grouping, should be applied to examine the quality of the model from a number of perspectives.

It should be mentioned that this is not the first application of sliding both the observational and model event identification threshold. As one example of this, in their presentation and initial usage of the extreme dependency score (EDS), Stephenson et al. (2008) simultaneously moved both thresholds. Events become rarer with increasing threshold and that study examined the relationship of EDS as a function of this rarity – moving both thresholds together, as is done here for the STONE curve.

A final note to make here is that this is not the first usage of the STONE curve. Both Liemohn, Ganushkina, et al. (2018) and Liemohn, McCollough, et al. (2018) used STONE curves in the plots labeled as ROC curves. It is clear that these panels are mislabeled because nonmonotonicities are seen in these lines.

## 5. Conclusions

A new data-model comparison assessment tool has been introduced, described, used, and interpreted – the sliding threshold of observations for numeric evaluation curve. Based on the relative operating characteristic curve, the STONE curve is created by plotting POD against POFD for a wide range of threshold settings. The main difference with the ROC curve is that the STONE curve requires the data to be continuous-valued real numbers and the model to be attempting to reproduce these exact values. The threshold is moved not only for the model, as is done for the ROC curve, but also for the observational event identification threshold setting, which is moved simultaneously with the model threshold setting.

The STONE curve has many features in common with the ROC curve with one large exception – it can have nonmonotonicities in both the POD and POFD values. For the ROC curve, the points shift within the quadrants defining POD or within the quadrants used to define POFD, but not between these two mutually exclusive regions. The ROC curve is, therefore, always monotonic, sweeping from (1,1) to (0,0) in POFD-POD space. For the STONE curve, the motion of the observational threshold moves points from the POD regions to the POFD regions, allowing for these nonmonotonic features in the STONE curve.

These wiggles and ripples, however, reveal information about the underlying distribution of points in the data-model scatter plot. Specifically, if the distribution is shifted, asymmetric, or bi-modal, the STONE curve will have a nonmonotonicity. Further investigation of the distribution, through a histogram, skew calculation, or other metric assessment, can reveal the true nature of the data-model comparison for this threshold setting.

It is hoped that the STONE curve becomes a useful data-model comparison tool. It has been used with two space weather applications in this study but these are purely illustrative examples. A dozen studies using ROC curves across the Earth and space sciences were given in the Introduction above. Some of these studies were based on observations that were pre-classified yes/no as events or not, and so the ROC curve is the proper tool for assessing the model's ability to predict those events. Some of those studies, however, and others like them, are based on models trying to exactly predict the observed data values, in which case the STONE curve might be a useful assessment tool. For any continuous-valued model trying to reproduce the exact numbers of a continuous-valued data set, the STONE curve can be calculated, perhaps, as shown for the two examples here, revealing additional information about the data-model comparison than can be obtained from the ROC curve alone. The STONE curve is a general purpose metric for use whenever a model is trying to exactly reproduce a continuous-valued data set. It can be used with both archival observations as well as for assessment of real-time nowcasting across the full breadth of science and engineering disciplines.

## Acknowledgments and Data

The authors would like to thank the US government for sponsoring this research, in particular research grants from NASA (NNX14AC02G, NNX16AG66G, NNX17AI48G, and NNX17AB87G) and NSF (AGS-1414517). The part of the research done by M. Liemohn and N. Ganushkina received funding from the European Union Horizon 2020 Research and Innovation programme under grant agreement 637302 PROGRESS. A. Azari's contributions are based on work supported by the NSF Graduate Research Fellowship Program (DGE 1256260). The SWMF simulations were conducted on the computing facilities at NASA GSFC and the run output is freely available on their website (https://ccmc.gsfc.nasa.gov/cgi-bin/SWMFpred.cgi) and the CCMC iSWA interactive tool (https://ccmc.gsfc.nasa.gov/iswa/). The real-time solar wind data was provided by NOAA SWPC (http://www.swpc.noaa.gov/products/real-time-solar-wind). The authors thank the World Data Center in Kyoto, Japan for the real-time Dst values (http://wdc.kugi.kyoto-u.ac.jp/dst_realtime/presentmonth/index.html). The IMPTAM simulations are available through the Finnish Meteorological Institute (http://imptam.fmi.fi/) and at the University of Michigan (http://citrine.engin.umich.edu/imptam/).

In addition to the archival repositories listed above, the specific observational data sets and the model output files used in this study are available at the University of Michigan Deep Blue Data repository, https://deepblue.lib.umich.edu/data/concern/data_sets/02870v99r?locale=en .